\documentclass[12pt,preprint]{aastex}

\shorttitle{3D Reconstruction of an Erupting Filament}
\shortauthors{ Li et al.}

\begin{document}

\title{Three-Dimensional Reconstruction of an Erupting Filament with SDO and STEREO Observations}

\author{Ting Li\altaffilmark{}, Jun Zhang\altaffilmark{}, Yuzong Zhang\altaffilmark{}, Shuhong Yang\altaffilmark{}}

\affil{Key Laboratory of Solar Activity, National Astronomical
Observatories, Chinese Academy of Sciences, Beijing 100012, China}

\email{[liting;zjun]@nao.cas.cn}

\begin{abstract}
On 2010 August 1, a global solar event was launched involving almost
the entire Earth$-$facing side of the Sun. This event mainly
consisted of a C3.2 flare, a polar crown filament eruption and two
Earth-directed coronal mass ejections (CMEs). The observations from
the Solar Dynamics Observatory (SDO) and the Solar Terrestrial
Relations Observatory (STEREO) showed that all the activities were
coupled together, suggesting a global character of the magnetic
eruption. We reconstruct the three-dimensional geometry of the polar
crown filament using observations from three different viewpoints
(STEREO A, B and SDO) for the first time. The filament undergoes two
eruption processes. Firstly, the main body of the filament rises up,
while it also moves towards the low-latitude region with a change in
inclination by $\sim$ 48$\degr$ and expands only in the altitudinal
and latitudinal direction in the field of view of Atmospheric
Imaging Assembly. We investigate the true velocities and
accelerations of different locations along the filament, and find
that the highest location always has the largest acceleration during
this eruption process. During the late phase of the first eruption,
part of the filament material separates from the eastern leg. This
material displays a projectile motion and moves towards the west at
a constant velocity of 141.8 km s$^{-1}$. This may imply that the
polar crown filament consists of at least two groups of magnetic
systems.

\end{abstract}

\keywords{Sun: coronal mass ejections (CMEs) --- Sun: filaments ---
Sun: flares}

\section{Introduction}

The twin Solar Terrestrial Relations Observatory (STEREO; Kaiser et
al. 2008; Howard et al. 2008) spacecraft provide simultaneous
observations from two different points of view, which enables us to
reconstruct the three-dimensional (3D) geometry of coronal
structures. After the launch of the STEREO mission, there have been
many studies about the 3D reconstructions using data from the pair
of Extreme UltraViolet Imagers (EUVI; see Wuelser et al. 2004). Feng
et al. (2007) presented the first reconstruction of the 3D shape of
coronal loops in an active region based on a triangulation
technique. Aschwanden et al. (2008a, b, 2009) discussed the 3D
geometry of coronal loops and determined the electron density and
temperature of loops. Besides the coronal loops, the STEREO data
have been used to reconstruct the active regions (Rodriguez et al.
2009; Wiegelmann \& Inhester 2006), polar plumes (Curdt et al. 2008;
Feng et al. 2009), polar coronal jets (Patsourakos et al. 2008) and
coronal mass ejections (CMEs; Mierla et al. 2008; Timothy \& Tappin
2008; Wood et al. 2009; Frazin et al. 2009).¡¡

Until now, there have been many studies about the nature of filament
eruption and its role in disrupting the coronal magnetic fields
(e.g., Zhang \& Wang 2001; Zhang et al. 2001; Liu et al. 2007; Jiang
et al. 2007; Liu et al. 2010; Guo et al. 2010a; Guo et al. 2010b).
The dynamics of erupting filaments are very important to understand
the associated physical mechanisms. However, filament velocity
derived using traditional method has a disadvantage because it is
only a projected velocity. Only recently, the true process of
filament eruption could be properly judged by using the STEREO data
(Li et al. 2010; Xu et al. 2010; Zapi\'{o}r \& Rudawy 2010; Gosain
et al. 2009; Liewer et al. 2009; Bemporad 2011; Panasenco et al.
2011; Thompson 2011). Gissot et al. (2008) presented a fully
automated optical-flow algorithm and estimated the height of an
erupting filament from simultaneously obtained EUVI images. Thompson
(2008) traced out a filament in 3D space and found a rotation of
$\sim$ 140$\degr$ from the original filament orientation. Bemporad
(2009) found that the early filament expansion was anisotropic and
occurred mainly on a plane parallel to the plane of the sky.

It is well known that the inner core of CME is made up of filament
material (e.g., House et al. 1981). The slow rise of eruptive
filaments linked to the streamer swelling is considered as the
precursor of CMEs (Filippov \& Koutchmy 2008). Thus an accurate
measurement of filament motion is very useful in predicting the
occurrence of a CME, as well as to constrain or construct the CME
triggering mechanisms (Gopalswamy et al. 2006). Nevertheless, only
several filaments have been reconstructed so far about their 3D
shape and evolution using the STEREO data. The 3D physical picture
of filament eruption is far from being completely understood. It is
disadvantageous to reconstruct a 3D configuration of solar features
by only using STEREO data because of the large separation angle
between the two satellites. With the launch of the Solar Dynamics
Observatory (SDO; Schwer et al. 2002), this would be much improved
by using data of SDO and STEREO. In this work, we make the 3D
reconstruction of a polar crown filament using observations from the
three different viewpoints, and use a visualization method to
display the reconstructed filament. Section 2 describes the
instrumentation and observations. The 3D reconstruction technique
and results are presented in Section 3. The eruptions of the polar
crown filament are described in Sections 4, and the conclusions and
discussion are given in Section 5.

\section{Instrumentation and observations}
The Atmospheric Imaging Assembly (AIA; Lemen et al. 2011) on the SDO
has already provided a rich database since its launch in February
2010. AIA images are taken in 10 different wavelength bands,
including one visible line, two ultraviolet and seven extreme
ultraviolet (EUV) channels. It provides full-disk images, covering a
wide range of temperatures, with high cadence (up to 12$-$24 s) and
spatial resolution (0.6 arcsec pixel$^{-1}$). The full-disk
longitudinal magnetic field from the Helioseismic and Magnetic
Imager (HMI; Schou et al. 2011) aboard SDO is available now, with
high cadence ($\sim$ 45 s) and spatial resolution (0.5 arcsec
pixel$^{-1}$).

The Sun Earth Connection Coronal and Heliospheric Investigation
(SECCHI; Howard et al. 2008) imaging package on each spacecraft of
the STEREO consists of the following five telescopes: the EUVI
imager, inner (COR1) and outer (COR2) coronagraphs, and inner (HI1)
and outer (HI2) heliospheric imagers. EUVI images are taken at four
wavelengths centered at 304 {\AA}, 171 {\AA}, 195 {\AA}, and 284
{\AA} with time cadences of 10, 2.5, 10, and 20 minutes,
respectively, for the data used here. COR1 has a field of view (FOV)
from 1.4 to 4 $R_{sun}$ and COR2 from 2.5 to 15 $R_{sun}$.

On 2010 August 1, a spectacular solar event was launched involving
almost the entire Earth$-$facing side of the Sun. At about 05:00 UT,
a large-scale ($\sim$ 750,000 km) polar crown filament stretching
across the solar northern hemisphere began to rise slowly. Then a
C3.2 flare occurred at 07:24 UT in active region 11092, with a
distance of $\sim$ 430,000 km from the long filament. At about 07:50
UT, an Earth-directed CME (CME front ``a" in Figure 1) was observed
by the COR1 aboard STEREO A. The average velocity of the first CME
in the FOV of COR1 was about 624 km s$^{-1}$, and the angular width
was $\sim$ 144$\degr$. At about 09:00 UT, another Earth-directed CME
occurred and the eruptive filament was the bright core of the second
CME (right panels in Figure 1). The second CME had an average
velocity of 260 km s$^{-1}$ in the FOV of COR1, and its angular
width was about 83$\degr$. Based on the Michelson Doppler Imager
(MDI) magnetograms, we carried out the global reconstruction of 3D
magnetic-field structure under the potential-field assumption. The
boundary element method (BEM) was used in the extrapolation (Yan \&
Sakurai 2000; Wang et al. 2002; Zhang et al. 2007). We found that
there were two groups of magnetic structures overlying the filament
(the large-scale magnetic structures denoted by character ``a" and
the smaller magnetic structures beneath the large-scale ones denoted
by ``b" in top panels of Figure 1).

Here, we concentrate on the eruption process of the large polar
crown filament. The filament eruption was recorded in all AIA
channels, and the 304 {\AA} images with a time cadence of 12 s were
used. The eruption was also observed as a limb event at 304 {\AA} by
the two EUVIs aboard STEREO. The pixel size of the EUVI image on the
solar disk was 1.6 arcsec, and the time cadence for 304 {\AA} images
was 10 minutes.

\section{Reconstruction technique and results}

For the first time, we combine data from SDO and STEREO to
reconstruct the erupting filament. The angle of separation between
STEREO A and B during our observations is around 149$\degr$.6. It is
difficult to define the same feature in EUVI A and B images because
of the large separation angle. The separation angle between STEREO A
and SDO is 78$\degr$.8, and that between STEREO B and SDO
70$\degr$.8. So observations from the three different viewpoints are
used here to derive the 3D erupting process of the polar crown
filament. The western part of the filament was observed clearly by
STEREO A and SDO, and thus we reconstruct the western part by using
data from STEREO A and SDO (Figures 2 and 3). Similarly, the eastern
part is reconstructed by using observations of STEREO B and SDO
(Figures 2 and 4).

In order to reconstruct the 3D shape of the eruptive filament, we
use a routine called SCC$_{-}$MEASURE (developed by W. Thompson),
which is available in the STEREO package of the Solar Software
library. The routine uses triangulation to determine the 3D
coordinate of the tiepointed feature. It is a widget based
application that allows the user to locate (and select with a
cursor) the same feature in two images from different views. This
routine makes use of an approximate ``epipolar constraint" in
locating the same feature in both images (Inhester 2006). The two
observer positions and any object point to be reconstructed exactly
define a plane, which is known as an epipolar plane. By definition,
epipolar planes are projected on both observer¡¯s images as lines
and these lines are called the epipolar lines. Once we identify a
feature in one image, it is possible to determine the projection of
the epipolar plane (i.e., epipolar line) passing the same feature in
the second image. After selecting the same feature in both images,
the 3D coordinates are then determined as longitude, latitude, and
radial distance from the center of the Sun.

It must be mentioned that there exists a limiting factor in
reconstructing filaments, because the reconstruction technique
relies on the ability to identify the same feature in two images
from different views. In fact, much of the corona is optically thin,
so the emission at each pixel may be the result of a line-of-sight
integration effect. Moreover, features may appear different from
different angles due to the projection effect. In order to reduce
the uncertainty, we examine the movies of filament eruption
carefully and determine the features according to their evolution.
We also use the bright or dark patterns in adjacent areas to
identify the features. As seen in Figure 7, the measurement is
repeated by two researchers, and each researcher measures twice. The
standard deviations in the four sets of measurements are considered
as the realistic uncertainties. The average values in the four sets
of measurements are thought to be the real values.


About four hours after the onset of the rise, the filament lifted up
to a high altitude above the solar surface and it became blurry
because of the steep radial intensity gradient. So we reconstruct
the 3D shape and evolution of the eruptive filament between 05:06
and 09:26 UT. We choose the top edge, the main-body line and the
bottom edge as three baselines of the filament. For the western part
of the filament, 20 pairs of points are placed along each baseline
in the images of STEREO A and SDO. For the eastern part, 16 pairs of
points are placed along each baseline in the images of STEREO B and
SDO. Therefore, 36 pairs of points are placed along each baseline
for the entire filament. We interpolate 500 points by using a cubic
spline along each baseline among the selected points to smooth the
baseline. Then we simply fill in the enclosed region between any two
baselines with 1000 triangular elements to produce the extended
regions. Using this visualization method, we can display the
reconstructed filament seen from any visual angle.


The reconstructed filament seen from the SDO viewpoint at three
different times is shown in Figure 2. The filament began to
gradually lift up at 05:06 UT. It spanned a longitude range of about
42$\degr$.3 (from 3$\degr$.3 E to 39$\degr$.0 W measured from the
central meridian) and a latitude range of about 26$\degr$.0 (from
20$\degr$.6 N to 46$\degr$.6 N). The filament had the initial length
of 1.07 $R_{sun}$, and the highest part at that time was
approximately at 1.15 $R_{sun}$. Then it moved towards the north
pole as seen from the SDO, and rose to a height of 1.31 $R_{sun}$ at
07:36 UT. At about 09:00 UT, the western part of the filament broke
up and then it was ejected from the Sun (bottom panels).

From the STEREO A perspective (Figure 3), the evolution of the
western part is displayed clearly. The filament has two main
endpoints at its west, composing a reverse$-$Y shape. The distance
between the two ends of the reverse$-$Y shape is 197.4 Mm. The
distance between the northern (southern) end and the intersection is
109.2 Mm (135.8 Mm). The northern endpoint could not be seen clearly
before the rapid ascent of the filament (top panels). With the
ascent of the filament, the reverse$-$Y structure is obviously seen
and the endpoint brightenings are observed simultaneously (bottom
panels). Strong and relatively concentrated brightenings occur at
the northern endpoint, while weaker and more dispersed EUV
enhancements are seen along the multiple, curved threads at its
southern endpoint. At 08:36 UT, the strands of the filament
anchoring the two ends are separated from each other. Then the
strands anchoring the northern end is disconnected, and the area of
brightenings at the northern endpoint expanded towards the
northeast. The barbs of the filament could be observed clearly from
the STEREO B perspective (denoted by white circles in Figure 4). At
05:06 UT, all the barbs are tightly anchored in the photosphere.
Then the barbs are elongated and finally cut off from the solar
surface in sequence due to the rise of the eastern part.

We reconstruct the eastern part of the filament by using data of
STEREO B and SDO, and reconstruct the western part using data of
STEREO A and SDO. For the eastern part, the STEREO A viewpoint is
the third view to confirm the result. The eastern part of the
reconstructed filament seen from STEREO A is similar to the observed
eastern part (Figure 3). Similarly, the STEREO B viewpoint is the
third view to confirm the result for the western part. The
reconstructed western part seen from STEREO B is similar to the
observed western part (Figure 4). These imply that our
reconstruction results are reliable to some extent.

We display the reconstructed filament seen from three different
points of view in Figure 5. The left panels show the results seen
from the top of the filament center, which is the central location
of the main body at 05:06 UT on the solar surface (denoted by plus
signs in panels (a) and (e); with a longitude of 12$\degr$.1 W and a
latitude of 39$\degr$.4 N). With the radial distance from the center
of the Sun increasing, the latitude of the filament decreases
gradually. The longitude of the filament remains approximately
invariable. This implies that the eruption is non-radial. The angle
between the radial direction (lines in panels (d)$-$(f); connecting
the filament center at 05:06 UT with the solar center) and the line
connecting the highest part with the filament center at 05:06 UT is
considered as the inclination of the filament (panel (e)). The
filament is initially inclined northward with an inclination of
$\sim$ 40$\degr$. Then it moves towards the low-latitude region and
is inclined southward with an inclination of $\sim$ 8$\degr$ at
09:26 UT. Seen from the north pole, the velocity of the filament has
a large Earth$-$directed component (panels (g)$-$(i)). This is
consistent with the Earth$-$directed CME associated with the
filament eruption (right panels in Figure 1).

\section{Eruptions of the filament}

\subsection{First eruption of the filament} In order to analyze the
filament eruption in detail, we investigate the true velocities and
accelerations of different sites along the filament. Eleven points
(points ``1" to ``11" in Figure 6) are selected along the filament
body at each time. Points ``1" to ``4" are defined using data of SDO
and STEREO B, and points ``5" to ``11" are selected in SDO and
STEREO A images. Due to motions of the material, it is difficult to
define the same feature during a period. So we examine the movies of
the filament eruption carefully, and select four specific points
(``1", ``5", ``10" and ``11") according to their continuous
evolution. Points ``1" and ``11" are near the two ends of the
filament. Points ``5" and ``10" are placed at the top and bottom
corners, respectively. Then we place other points equally between
each two specific neighboring points according to SDO/AIA images.
Thus all the 11 points along the main body are defined at each time.

Because the velocity-location plots of the 11 points at different
times are similar, we only display the plots at three times (Figures
7(a) and (b)). Point ``5" at the top corner of the filament is
always the highest among the 11 points. It also has the largest
velocity and acceleration at any time. Moreover, the true heights,
velocities and accelerations of the highest point between 04:56 UT
and 09:36 UT are presented (Figure 7(c)). The height of the filament
increased by 0.72 $R_{sun}$ over four hours. There is some evidence
for a slight, slow rise before the fast eruption. Before 07:06 UT,
the filament lifted up gradually; the velocity and acceleration were
less than 13.8 km s$^{-1}$ and 3.0 m s$^{-2}$, respectively. Then it
ascended more rapidly and reached a velocity of 144.8 km s$^{-1}$ at
09:26 UT. Arrows in Figure 7(c) show the time of the C3.2 flare
start (07:24 UT) and peak (08:56 UT). About 30 minutes before the
flare onset, the acceleration of the highest point began to increase
and reached a value of 9.2 m s$^{-2}$ at 07:36 UT. Then it decreased
to 5.2 m s$^{-2}$ at 07:56 UT and increased to 41.2 m s$^{-2}$ at
09:16 UT afterwards. When the velocity of the filament continuously
increased to about 90 km s$^{-1}$ at 09:06 UT, the associated CME
was clearly observed in STEREO A/COR1 images. At this time, the
velocity of the CME was 163 km s$^{-1}$ (Figure 7(d)), which is more
than the velocity of the filament by a factor as large as 1.8. Then
the velocity of the second CME increased to 762 km s$^{-1}$ at 13:18
UT in the FOV of COR2 (Figure 7(d)).


We also investigate the variations of the longitudes and latitudes
for the reconstructed features during the fast-eruptive phase
(between 07:06 and 09:26 UT). As seen in Figure 8, the longitudes of
points ``2" to ``4" at the eastern leg show a decrease of $\sim$
7$\degr$, while points ``6" to ``9" at the western leg show an
increase in longitude of about 12$\degr$. The latitudes of all the
points except ``9" are seen to decrease by $\sim$ 5$\degr$. The
measurement is repeated four times, and the corresponding errors in
longitude and latitude are about 0$\degr$.6 and 0$\degr$.3,
respectively. Similar to the analysis of Joshi \& Srivastava (2011),
our analysis indicates that the filament experiences a twist in a
counterclockwise direction.

In the process of the filament eruption, the EUV enhancements at the
endpoints are observed clearly (top panel in Figure 9). At about
07:20 UT, the initial brightenings occurred at the eastern
endpoints. At 08:20 UT, the brightenings at the western endpoints
were observed. Comparison with the magnetograms shows that the
western (eastern) ends of the filament are rooted in negative-
(positive-) polarity fields (Figure 9). The magnetic polarity of the
ends determines the direction of the axial field component of the
filament. According to the empirical rule of Martin (1998), if the
barbs point forward and to the right (left), then the axial field
points to the right (left) when viewed from the positive-polarity
side of the polarity inversion line, and the filament has
``left-handed" (``right-handed") chirality. Therefore, the filament
is left-handed.

\subsection{Second eruption of the filament}

At the late phase of the first eruption ($\sim$ 09:30 UT), the
eastern leg of the erupting filament obviously started to expand.
Then partial material from the filament separated from the eastern
leg (see rectangle ``1" in Figure 10(b)). The separated material
gradually diffused towards the west and the orientation of the
separated material became mainly northeast-southwest at 10:16 UT
from the longitudinal direction at 09:51 UT (Figure 10(b)). This
material displayed a projectile motion and moved along the filament
channel in AIA images (Figure 10(a)). The material became invisible
at 11:25 UT when it moved to the west limb.

In order to follow the motion of the material in detail, we select
the slice (line A-B in Figure 10) along which the material moves
from the running ratio 304 {\AA} images. As seen in the running
ratio stack plot (Figure 10(c)), the material moves at a nearly
constant speed. The velocities of two clear structures are about
145.2 km s$^{-1}$ and 138.3 km s$^{-1}$. In the first eruption, the
main body of the filament lifted up and was accelerated (Figure
7(c)). In the second eruption, partial material of the eastern leg
displayed a projectile motion toward the west and the velocity
remained constant.


\section{Conclusions and discussion}

On 2010 August 1, a C3.2 flare, a polar crown filament eruption and
two Earth-directed CMEs were coupled together, indicating the global
character of solar activity (Wang et al. 2007; Zhukov \& Veselovsky
2007; Zhou et al. 2007). Schrijver \& Title (2011) analyzed a series
of events occurring on 2010 August 1-2, and concluded that all
substantial coronal activities were initiated from a connected
network of large-scale separators, separatrices, and
quasi-separatrix layers. Also they found that the magnetic field
lines emanating from the flare region slid over a quasi-separatrix
layer connecting the polar crown filament.

For the first time, we reconstruct the polar crown filament using
observations from three different viewpoints (STEREO A, B and SDO).
The initial length of the filament was 1.07 $R_{sun}$, and the
highest part was approximately at 1.15 $R_{sun}$. Three hours after
the initial ascent, it rose to a height of 1.47 $R_{sun}$ and the
length increased to 1.75 $R_{sun}$, about 1.6 times as long as the
initial length. Two western endpoints and a partial spine of the
filament compose a reverse$-$Y structure. During the eruption, the
barbs are elongated and finally cut off from the solar surface in
sequence. AIA observations show that the slow-rise phase of the
filament lasts about two hours, and the velocity and acceleration at
the highest location are less than 13.8 km s$^{-1}$ and 3.0 m
s$^{-2}$, respectively. Then it ascends more rapidly and reaches a
velocity of 144.8 km s$^{-1}$ at 09:26 UT and an acceleration of
41.2 m s$^{-2}$ at 09:16 UT. The highest location always has the
largest acceleration, implying that the location bears the largest
force during the eruption process. This result is different from a
former study (Li et al. 2010) that the location bearing the largest
force varies during a filament eruption process.

The reconstructed filament is processed with a visualization method
and we can view it from any visual angle. The reconstructed western
part of the filament gained with STEREO A and SDO data seen from the
third viewpoint (STEREO B) is similar to the observations. This
indicates that our reconstruction results are reliable to some
extent. Seen from the north pole, the velocity of the filament has a
large Earth$-$directed component. This is consistent with the
Earth$-$directed CME accompanying the eruptive filament. The
filament moves towards the low-latitude region, with an inclination
change of $\sim$ 48$\degr$ with respect to the radial direction in
the FOV of AIA. The filament expands only in altitude and in the
latitudinal direction, similar to the conclusion made by Bemporad
(2009), who found that the radial and latitudinal expansions of the
filament are much larger than the longitudinal ones and suggested
that the filament can be approximated as a 2D ``ribbon-like"
feature, instead of a 3D twisted flux tube.

In addition, at the late phase of the first eruption, partial
material separates from the eastern leg. This material displays a
projectile motion, and moves towards the west at a constant velocity
of 141.8 km s$^{-1}$. To our knowledge, this phenomenon has never
been reported before. The velocity of the second eruption remains
approximately constant, while the velocity of the first eruption
increases with time, which indicates that their eruption mechanisms
are different. Hirose et al. (2001) concluded that the filament with
helical fields accelerate upward in a round-shaped expanding field
structure forming a CME containing the eruptive filament. The first
eruption corresponds to the rise of helical magnetic fields of the
filament (Jing et al. 2004; Sterling et al. 2007; DeVore \&
Antiochos 2008), and the second eruption is the motion of filament
material along the magnetic system due to unequal magnetic pressure.
The observations may imply that the polar crown filament consists of
at least two groups of magnetic systems.

\acknowledgments {We are grateful to Dr. William. T. Thompson and
Dr. H. Zhao for useful discussion. We acknowledge the SECCHI, AIA
and HMI consortia for providing the data. This work is supported by
the National Natural Science Foundations of China (40890161,
11025315, 10921303 and 11003026), the CAS Project KJCX2-YW-T04, the
National Basic Research Program of China under grant 2011CB811403,
and the Young Researcher Grant of National Astronomical
Observatories, Chinese Academy of Sciences.}

{}

\clearpage

\begin{figure} \centering
\includegraphics
[bb=79 106 506 736,clip,angle=0,scale=0.8]{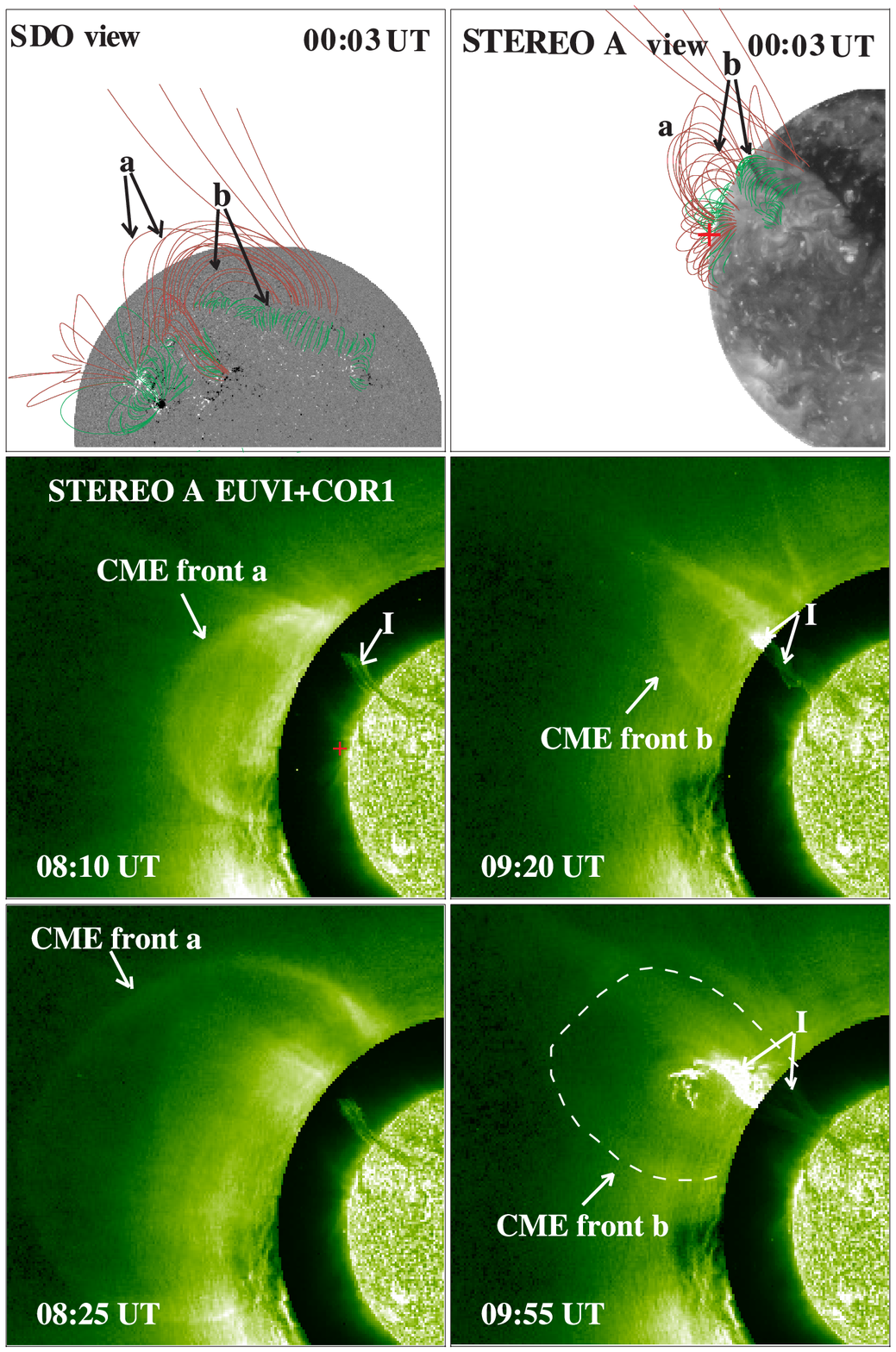}
\caption{Extrapolated coronal magnetic field lines seen from the SDO
(top left) and STEREO A (top right) perspectives at 00:03 UT on 2010
August 1 and composite images (middle and bottom panels) of STEREO-A
COR1 and EUVI 304 {\AA}. The backgrounds in the top panels are the
MDI line-of-sight magnetogram at 00:03 UT and the STEREO A 195 {\AA}
image at 00:05 UT, respectively. Characters ``a" and ``b" denote the
large-scale and small magnetic structures. Red and green curves
represent high and low magnetic lines. Arrows ``I" indicate the
filament material. The plus signs denote the flare
position.\label{fig}}
\end{figure}
%
%


\begin{figure}
\centering
\includegraphics
[bb=27 167 565 694,clip,angle=0,scale=0.8]{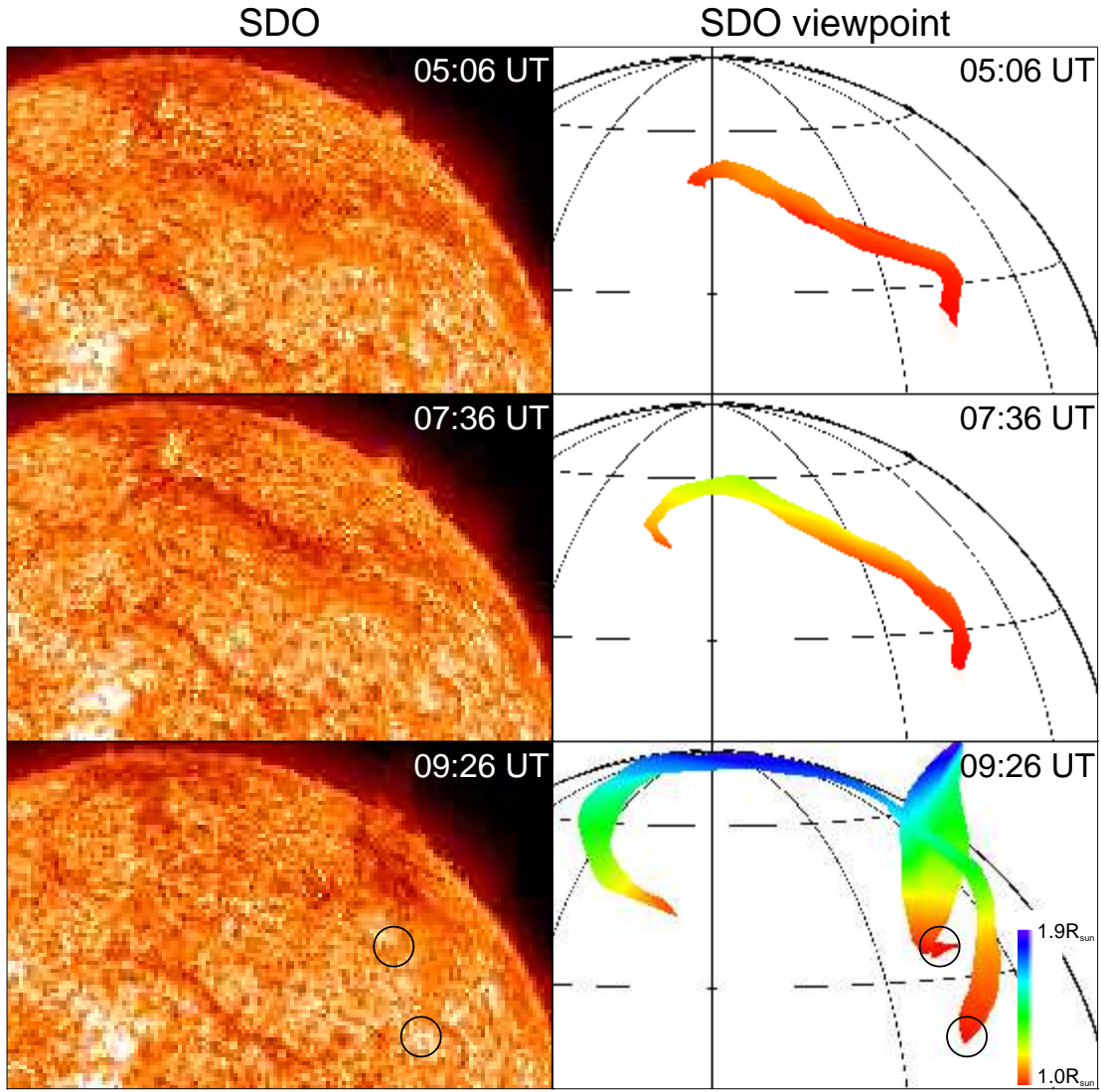}
\caption{Reconstructed filament seen from the SDO viewpoint. Left: a
series of SDO/AIA 304 {\AA} observed images showing the evolution of
the filament. Right: a series of reconstructed images seen from the
SDO viewpoint showing the eruption process of the filament. The
black circles denote the western endpoints of the
filament.\label{fig}}
\end{figure}

\begin{figure}
\centering
\includegraphics
[bb=120 137 476 723,clip,angle=0,scale=0.8]{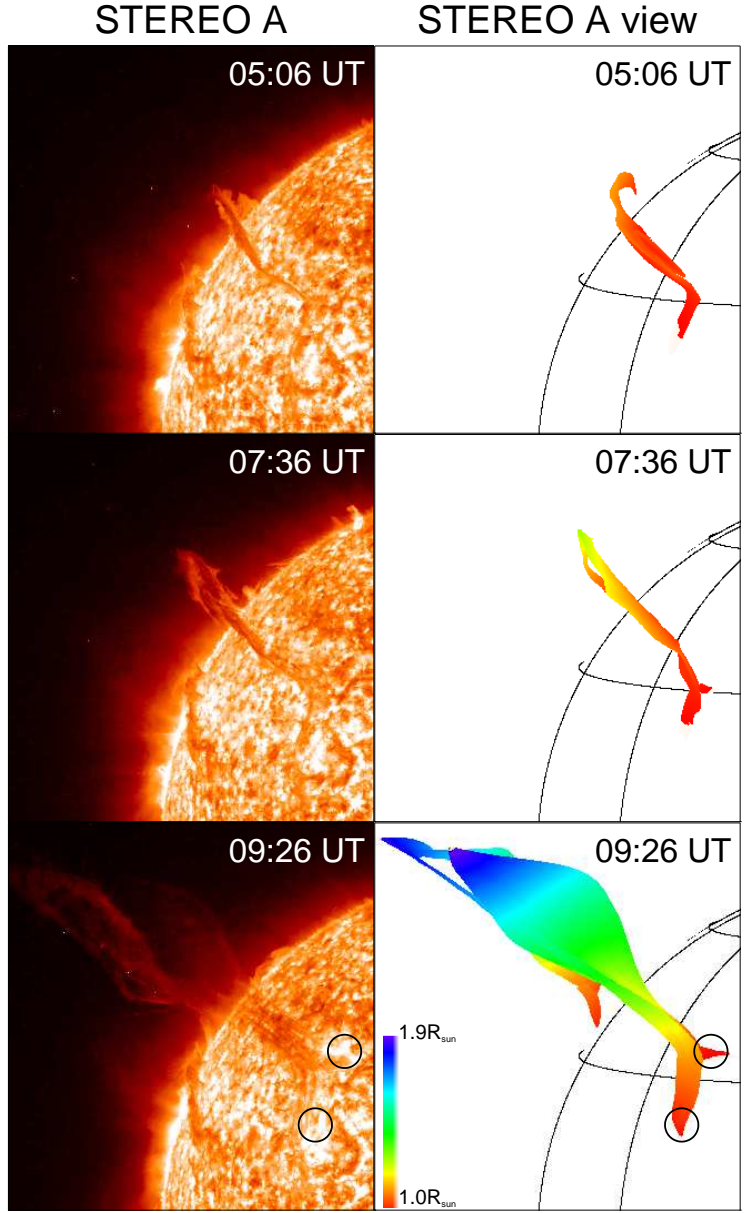}
\caption{Comparison of the observed filament (left panels) and the
reconstructed filament (right panels) seen from the STEREO A view.
The black circles denote the western endpoints of the
filament.\label{fig}}
\end{figure}
\clearpage

\begin{figure}
\centering
\includegraphics
[bb=120 165 477 687,clip,angle=0,scale=0.8]{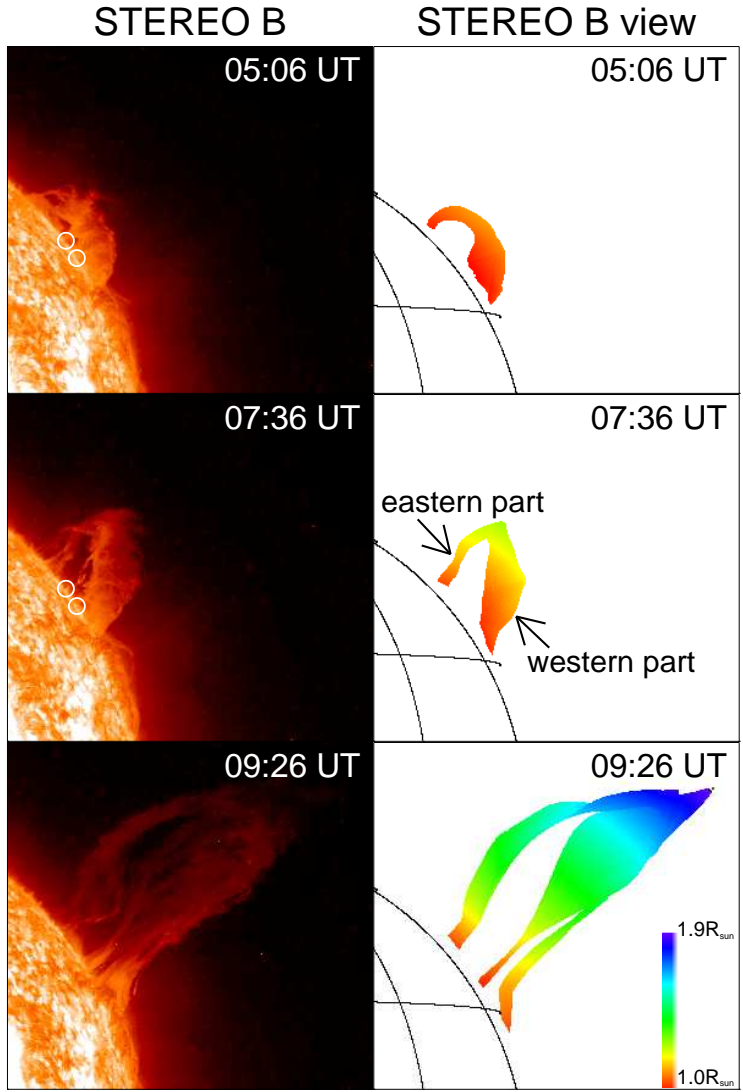}
\caption{Comparison of the observed filament (left panels) and the
reconstructed filament (right panels) seen from the STEREO B view.
The white circles denote the barbs of the filament.\label{fig}}
\end{figure}
\clearpage

\begin{figure}
\centering
\includegraphics
[bb=55 178 537 671,clip,angle=0,scale=0.8]{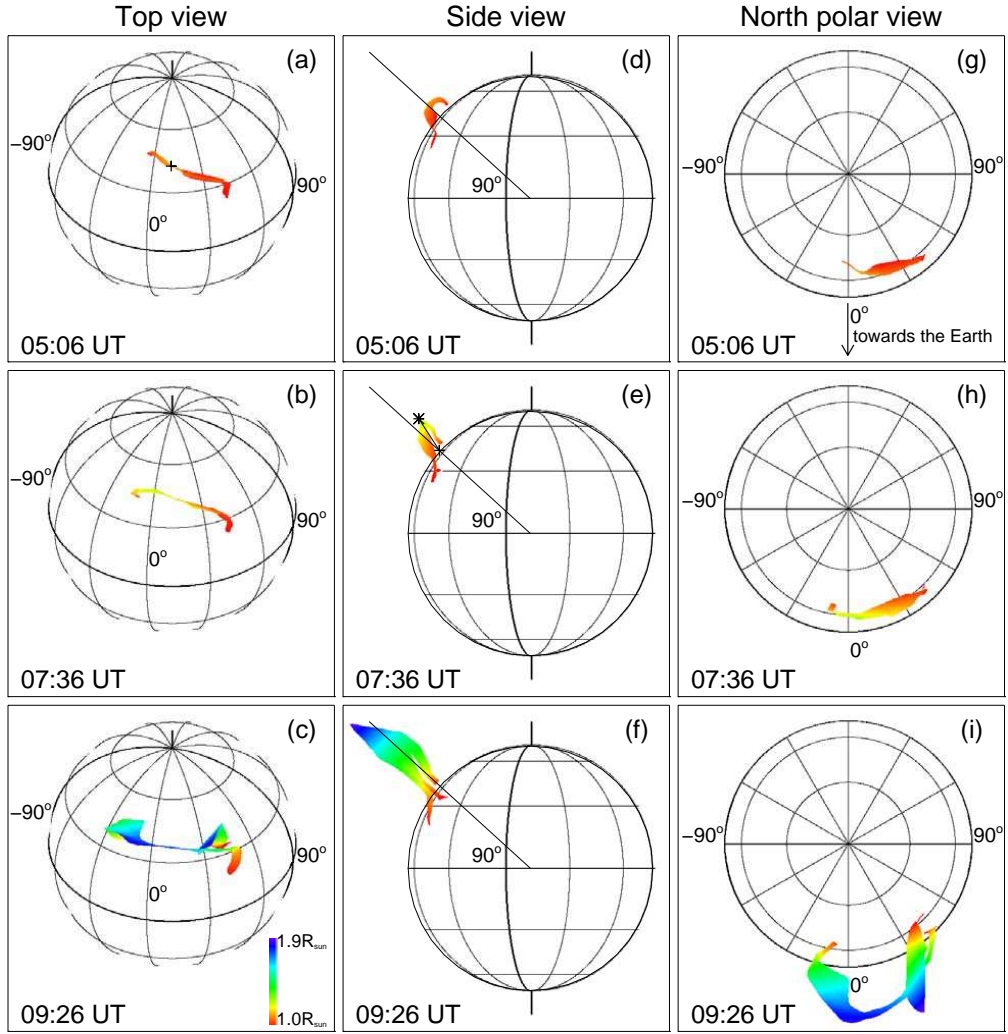}
\caption{Evolution of the reconstructed filament seen from the top
view (left panels), the side view (middle panels) and the north
polar view (right panels). In the heliographic coordinate system,
the top view of the filament has a longitude of 12$\degr$ and a
latitude of 39$\degr$, and the side view has a longitude of
102$\degr$ and a latitude of 0$\degr$. \label{fig}}
\end{figure}
\clearpage

\begin{figure}
\centering
\includegraphics
[bb=26 120 585 720,clip,angle=0,scale=0.8]{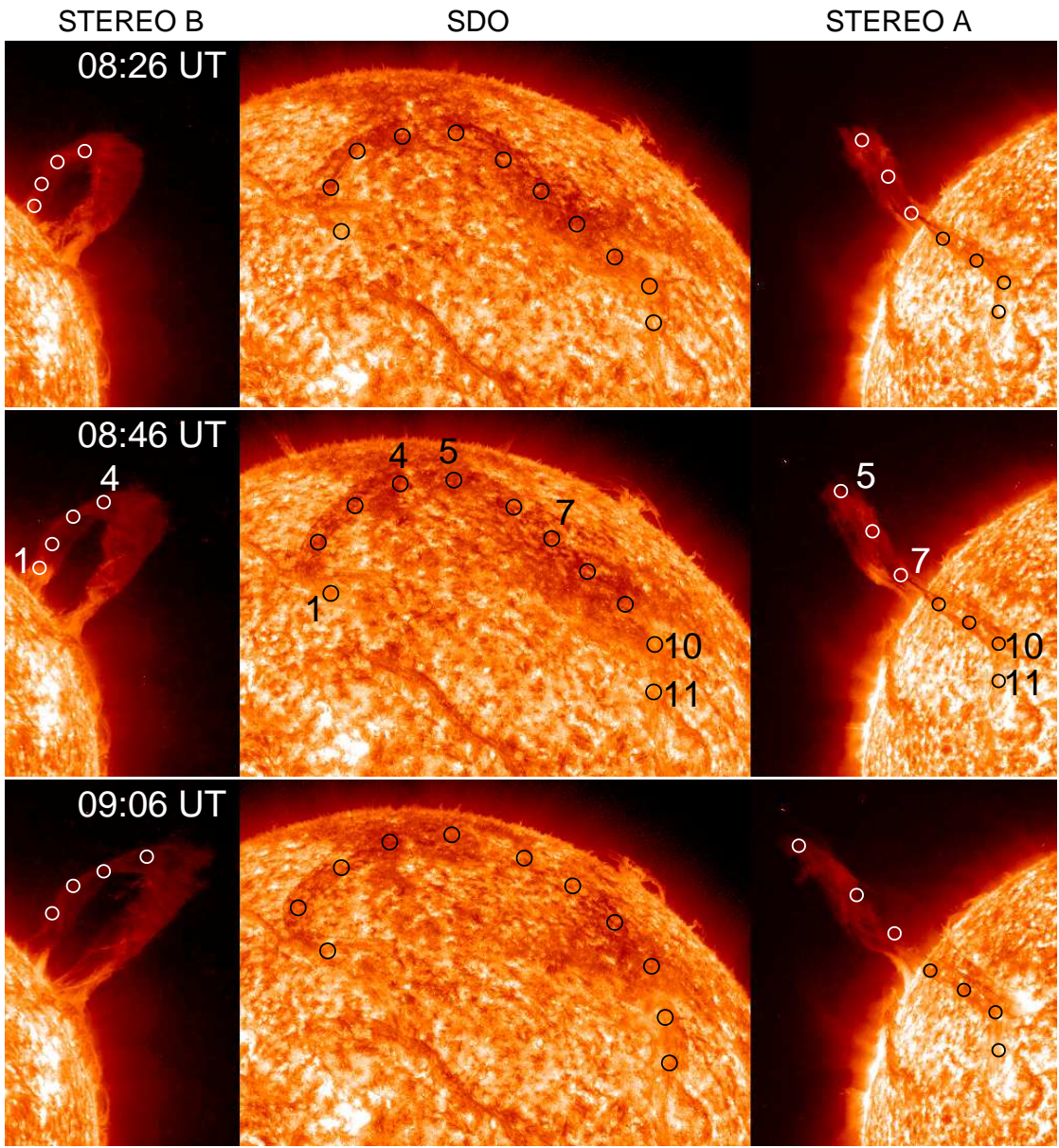}
\caption{Locations of 11 points selected in SDO and STEREO images.
The circles denote the 11 points along the filament. \label{fig}}
\end{figure}
\clearpage

\begin{figure}
\centering
\includegraphics
[bb=90 128 560 682,clip,angle=0,scale=0.8]{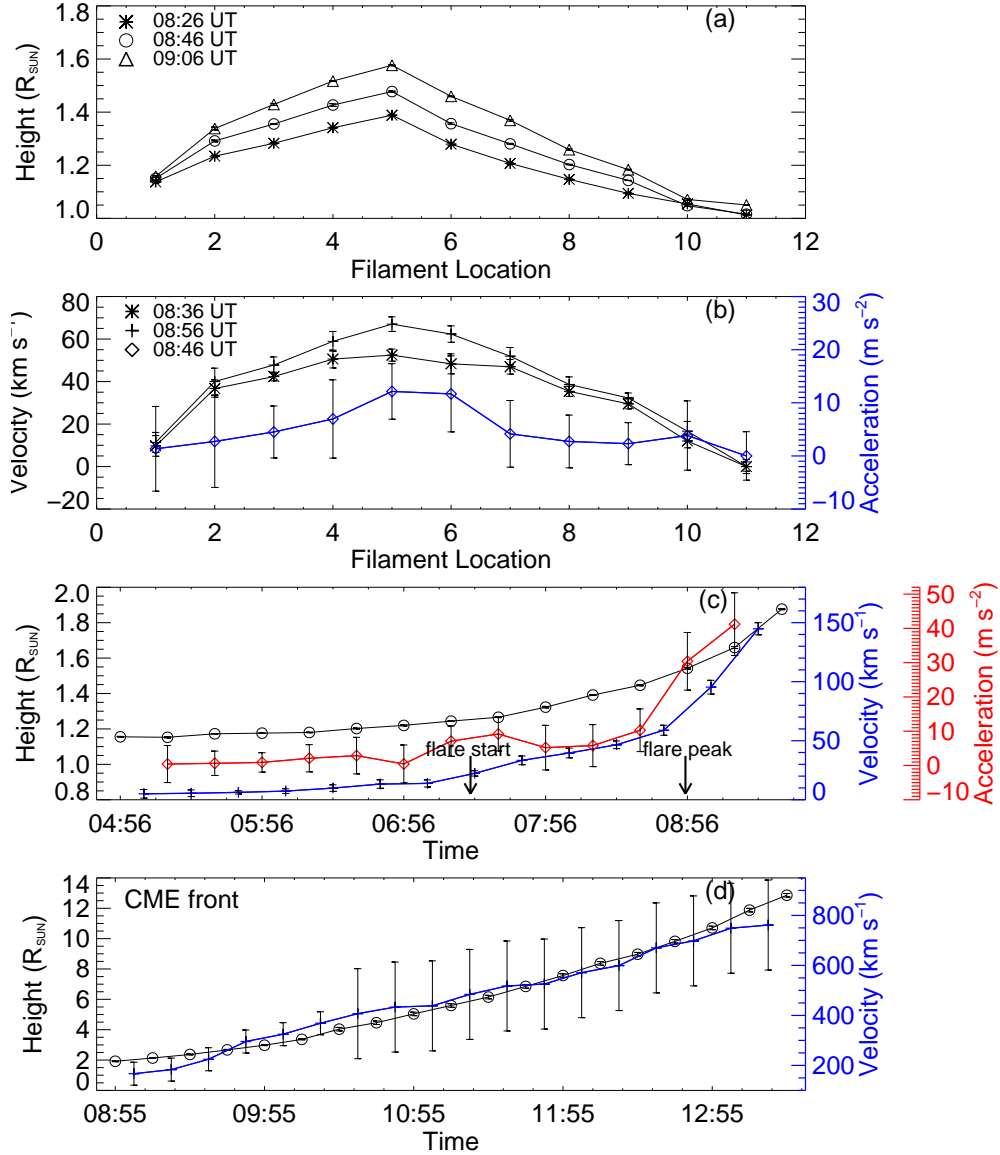} \caption{(a):
True height-location plots of the 11 points (points ``1" to ``11" in
Figure 6) at three different times. (b): Velocity-location and
acceleration-location plots of the 11 points. (c): True height-time,
velocity-time and acceleration-time plots of the highest point
between 04:56 UT and 09:36 UT. (d): Projected height-time and
velocity-time plots of the associated CME measured in STEREO A/COR1
and COR2 images. \label{fig}}
\end{figure}

\begin{figure}
\centering
\includegraphics
[bb=85 231 463 574,clip,angle=0,scale=0.8]{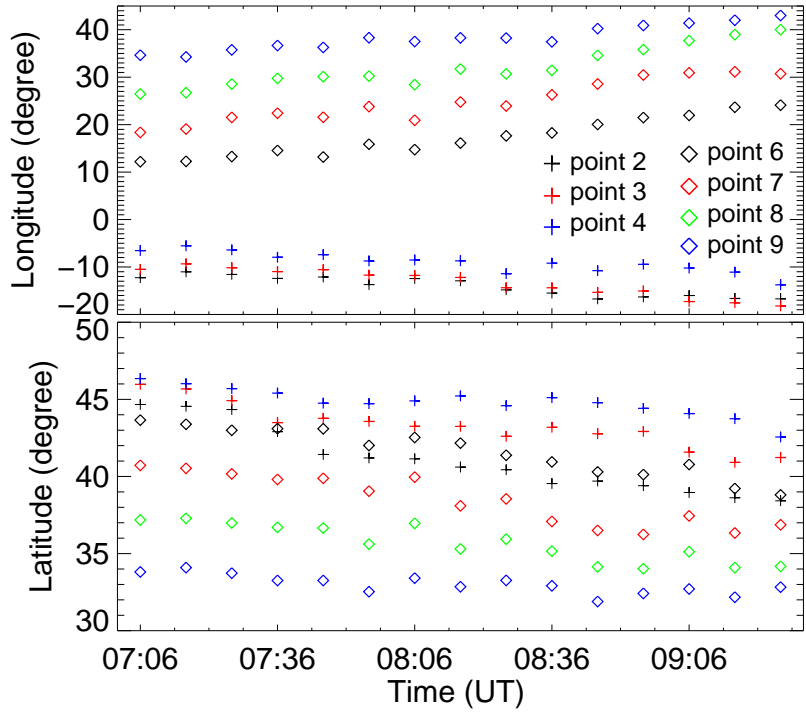}
\caption{Longitude-time and latitude-time plots of different points
of the filament. Points ``2" to ``4" are from the eastern leg, and
points ``6" to ``9" are from the western leg (Figure 6).
\label{fig}}
\end{figure}

\begin{figure}
\centering
\includegraphics
[bb=73 168 518 672,clip,angle=0,scale=0.8]{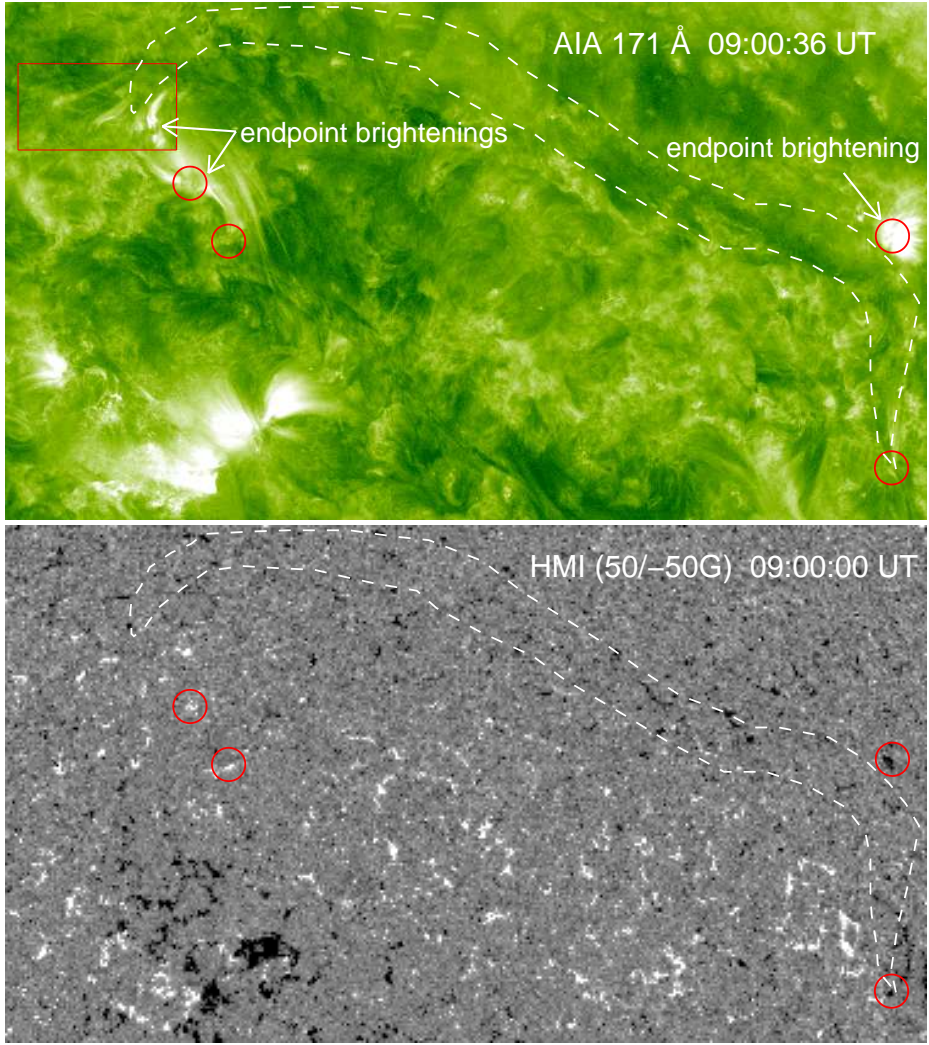} \caption{AIA
171 {\AA} image (upper panel) and HMI line-of-sight magnetogram
(lower panel) at the filament location. The white contours are the
filament at 05:06 UT. The red circles denote the multiple endpoints
of the filament. The red rectangle is the 171 {\AA} image at 07:40
UT. \label{fig}}
\end{figure}
\clearpage


\begin{figure}
\centering
\includegraphics
[bb=83 87 454 720,clip,angle=0,scale=0.7]{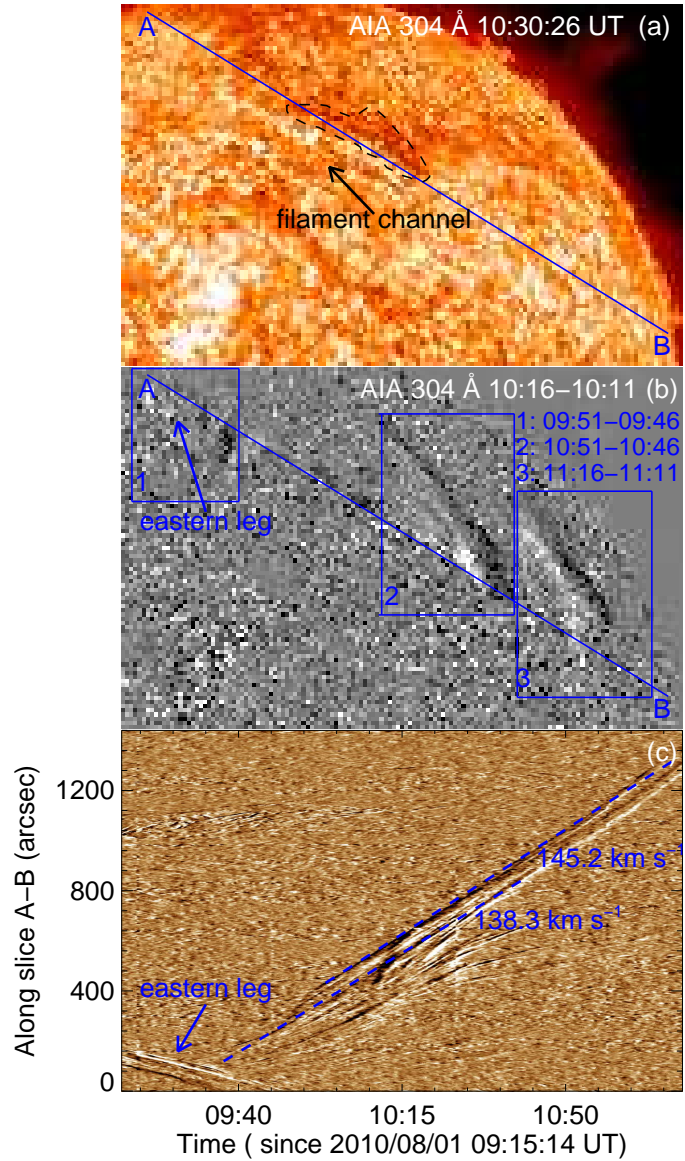}
\caption{Appearance of the second eruption of filament material seen
in AIA 304 {\AA} (panel (a)) and running-difference images (panel
(b)) and the running ratio stack plot obtained along slice A-B
(panel (c)). The black contour denotes the filament material along
line A-B.\label{fig}}
\end{figure}
\clearpage


\end{document}